\documentstyle{article}
\begin{document}

\title{Computation of Structure Functions\\
 From a Lattice Hamiltonian}

\author{N. Scheu\\
Universit\'e Laval\\
D\'epartement de physique\\
Qu\'ebec, Qu\'ebec, G1K 7P4 \\
}

\date{ }
\maketitle

\begin{abstract}

We suggest to compute structure functions in the Hamiltonian formalism
on a momentum lattice using a physically motivated regularisation that links
the total parton number to the lattice size.
We show for the $\phi ^4 _4$ theory that our method allows to describe
continuum physics. The critical line and the renormalised mass spectrum close
to that critical line are computed and scaling behaviour is observed
in good agreement with
the semi-analytical results of L{\"u}scher and Weisz and with other
lattice simulations.
We also demonstrate
that our method is able to reproduce the $Q^2$ behaviour of
deep inelastic structure functions and the typical peak at $x_B=0.$

\end{abstract}
\section{Introduction}

Hadron structure is probed by deep inelastic scattering (DIS). Much
effort has been devoted by theorists to
compute quark or gluon distribution
functions and proton structure functions from QCD with non
perturbative methods. E.g., Martinelli
et al.\cite{Martinelli} have computed the first two moments of the
pion structure function. These calculations are notoriously difficult; for
the present status of lattice calculations based on the Euclidean
action formulation see Ref.\cite{Lattice}.
In order to compute structure functions it seems natural to use the Hamiltonian
formulation that allows to compute wave functions. E.g., scattering wave
functions for glueball-like states in compact $QED_{2+1}$ have been computed
in a Hamiltonian formulation on a momentum
lattice\cite{ChaaraHelmut}. For a review of Hamiltonian lattice methods
see\cite{HelmutsReview}\cite{Schutte}.
The usefulness of a momentum lattice
to compute physics close to a critical point  has
been demonstrated in Ref.\cite{HelmutStephane}\cite{Espriu}
for the $\phi^4$-model.
Unfortunately, Hamiltonian methods are known to pose problems
when calculating the
huge number of field degrees of freedom due to particle number.
Nobody has yet succeeded in seeing scaling
laws indicating continuum physics in a $QCD$-Hamiltonian formulation,
to the best of the authors' knowledge.
In this work we shall demonstrate that these difficulties can be
overcome.
We introduce a new type of regularisation which excludes all
parton momenta that we deem to be not directly related to
deep inelastic structure measurements. It turns out that such a regularisation
scheme provides
both an ultraviolet as well as a particle cutoff.
We have chosen to test our method on the
popular scalar $\phi^4$ theory in {\it four} space-time dimensions because
this model has been thoroughly investigated numerically and
analytically.
The principal results are:

1. Using a Hamiltonian approach we observe continuum physics  (scaling
laws close to the critical line) for the $\phi^4 _4$ and we see

2. Altarelli-Parisi-type behaviour of $\phi ^3$ distribution
functions in two, three and four dimensions and a peak at $x_B=0$ similar
to the peak occurring in quark distribution
functions.

\section{DIS and a physical choice of Fock space}

Physical observables of relativistic
field theories depend continuously
on the experimental resolution.
Every experiment has a
finite resolution and coarse-grains the observed system. This corresponds
{\it cum grano salis} to
a regularisation of the field theory which describes
the observed system. The hadron structure functions
that we attempt to calculate are a prominent example.
The most important experiment in order to probe the structure of hadrons
is deep inelastic scattering (DIS). Its simplest form is inclusive
scattering of an
unpolarised lepton off a hadronic target which we want to build
our method upon.
For the basics of DIS see, e.g.\cite{Roberts}. Notation:
The hadron in its ground state with four momentum $P$ interacts with
the probing lepton by the exchange of a virtual photon (our neutrino)
with space-like four-momentum $q$.
In Feynman's parton
model, the impulse approximation is invoked. It is assumed that the proton
consists of constituents,
the partons, that are weakly bound when compared to the resolution ability
$Q:=\sqrt{-q_\mu q^\mu}$ of the probing photon.
In this approximation, the so-called Bj{\o}rken scaling variable
 $x_B:= {Q^2\over 2P_\mu q^\mu}$ can be
interpreted as the momentum fraction of the struck parton in the Breit frame
(which is defined by the requirements that the photon energy $q_0$ be zero and
that the momentum $\vec q$ point in the opposite direction of the hadron
momentum $\vec P$). In this model, structure functions $F(x_B,Q)$ can be
interpreted as some linear combinations of parton momentum distribution
functions $f(x_B,Q)$. For details see\cite{Roberts}.

We want to design a theory where just this picture emerges, describing only
the coarse-grained particles that are probed in a DIS experiment.
Consequently, we have to devise a regularisation that mimics the
coarse-graining due to experiment as close as possible.
In order to achieve this goal we have to decide on a conveniant Lorentz frame,
because firstly the Hamiltonian
is only defined with respect to such a frame and, secondly, a parton being a
virtual
particle is only defined with respect to the specific relativistic frame
where experiment performs its coarse-graining of space.
Seen from a different Lorentz frame, the parton would emerge as a jet of many
partons because the boost generators of the Poincar\'e group contain vertices
that change the parton number!
Our candidate is the Breit frame, of course. The struck parton in the Breit
frame,
however,
is collinear to the hadron momentum.
Thus, we have to allow for non-collinear partons
in more than one space dimension.
Therefore we use a {\it 'generalised Breit frame'} where $q_0=0$ but $\vec q$
does {\it not} have to be collinear to $\vec P$. Please note that only in such
a
'generalised Breit frame' is
$Q=|\vec q|$ and only there can the wave length
$\lambda= {2\pi\over|\vec q|}$ of the virtual photon be interpreted as the
resolution ability of the experiment! Please note also that
the momentum $\vec P$ strongly restricts the domain of all
possible photon momenta $\vec q$ in the generalised Breit frame.
We require therefore that the
{\it 'observable' Fock space} consist only
of partons whose momenta can (but need not) reverse momentum
$-\vec p\equiv \vec p
+\vec q$ if struck by a virtual
photon in the generalised Breit frame. Parton momenta that can not be
related by $\vec p=-\vec q/2$ to some virtual photon with momentum $\vec q$
are skipped.
Then all
 'observable' momenta {\it of the ground state} lie inside a sphere

\begin{equation} \label{criterion}
(\vec p-\vec P/2)^2 \leq |\vec P/2|^2
\end{equation}

around $\vec P /2$ with diameter $|\vec P|$  because $x_B={Q^2\over 2Pq}=
-{Q^2/2\over \vec P \cdot \vec q}$ lies between 0 and 1.

If we go onto a momentum space lattice with
momenta $\vec k:= \vec n \Delta k$ where $\vec n$ is an integer vector
and ${\Delta k}$ is the momentum lattice spacing, then the constraint
(\ref{criterion}) on the parton momenta
provides an ultraviolet as well as a particle cutoff. This holds because only
a finite number of partons with {\it positive} longitudinal momentum
$p_e:=\vec p\cdot \vec e$ can share the {\it positive} total longitudinal
momentum $\vec P\cdot\vec e$, where $\vec e:=\vec P/ |\vec P| $ denotes
the direction of the total hadron. If one performs the continuum limit,
the system's velocity $\vec v=\vec P/E$ grows and one ends up with
the velocity of light. For a parton with momentum fraction $x_B=const$ the
continuum limit $a\rightarrow 0$
also implies $Q= {aQ\over a}\rightarrow\infty$ and hence
is equivalent to the Bj{\o}rken limit. By the way, the transverse momenta
in this limit are
suppressed by the increasingly dominant kinetic term, at least in
asymptotically free and trivial
theories.

Since we want to calculate the critical line -- which depends on the
regularisation -- we had to ensure that our regularisation(\ref{criterion})
be well comparable to to the lattice regularisation of ref.\cite{Luscher}.
To this aim we started on the usual lattice with the position space
lattice spacing $a={\pi\over N\Delta k}$ as ultraviolet regulator,
where $2N+1$ denotes the number
of lattice points in each space direction, and put the total momentum $\vec P$
on one of the edges $\vec P=(1,1,1)\Delta k N$ of the momentum lattice.
If we then apply our regularisation(\ref{criterion}), every component
of the parton momenta is positive, i.e the combination of our
regularisation with the lattice cutoff restricts the parton momenta
to the first octant of the momentum lattice which further simplifies the
set-up of the fock space. The maximal number of 'partons' equals $N$
in this description. We ought to add that for a small lattice
such as $2N+1=7$ the additional lattice spacing cutoff does not exclude
states that are allowed by our regularisation -- so this seems to be o.k..

A further feature of the proposed ansatz is remarkable: the connection between
 boosts
and renormalisation group transformations. A boost that reduces the total
momentum $|\vec P|$
equally reduces the number of observable Fock states in our approach and hence
constitutes
a coarse-graining procedure on a momentum grid with fixed
infrared regularisation-- in contrast to that, a
boost without our regularisation would
increase the (unobservable) particle number.
Since the hadron momentum $|\vec P|={\pi \over a}$ is the cutoff in
our approach, the scale
$1/a$ is reduced and the bare parameters have to be re-adjusted. This
means that such a boost acts in a way similar to a renormalisation
group transformation in the Kadanoff/Wilson sense. Such a transformation
is not invertible.
Intuitively this corresponds to the fact that decreasing the momentum
of the probing lepton relative to the target hadron corresponds to a
lesser structure resolution.
If $|\vec P|$ gets too small we have coarse-grained too much and the
system breaks down, of course. In the momentum-sector with $\vec P=\vec 0$
only one state, the vacuum, remains.
Within our method we cannot describe a hadron in its rest frame, just
as one cannot perform structure measurements without a particle beam moving
relative to the target.
This means that we cannot give a Fock space expansion of the
physical vacuum either. This is no draw-back,
it is a desirable feature.
It makes simply no sense to calculate structure functions for the vacuum.
We treat it as a background field with the same quantum numbers
as the perturbative vacuum because only vacuum expectation values
of fields or of the energy (Casimir effect) originate from the vacuum.
The fact that the perturbative vacuum is annihilated by our Hamiltonian
does not mean that it is the physical vacuum --the latter one is far
from being trivial. It only means that per constructionem our theory
is designed for deep inelastic scattering and breaks
down if we set $\vec P$ equal zero.
If we, however, boost the momentum grid also, the transformation
is an invertible scale transformation making the physical quantities change
according to the Callan-Symanzik equation.
Three little remarks at the end of this section:

1. Theoretically, our regularisation(\ref{criterion}) admits partons with zero
momentum.
An infinite number of these can be present because they do not
take away longitudinal momentum. It is these that are responsible for
vacuum expectation values. Fortunately, for a system with
{\it spontaneous} symmetry breaking as the $\phi^4$ theory,
we can treat them on the {\it classical} level since at least in the infinite
volume limit, quantum mechanical fluctuations of the vacuum expectation values
must vanish. To describe the axial anomaly of a non-abilian gauge theories,
however, we shall have to allow for dynamic
gauge fields with zero momentum in order to describe fluctuations of the
Dirac sea.

2. It is obvious that our approach sheds some light on the relation of the
so-called
front- and instant forms\cite{BrodskyPauli} in two dimensions.
A discussion of this must be reserved for another paper.

3. Our method would, of course, permit to calculate the hadronic tensor
 and thus the structure functions directly from the electromagnetic
courrent operators.
But the calculation of distribution
functions is much easier.

\section{Application on the scalar $\phi^4 _4$ theory}

Now we apply our method to the $\phi^4_4$ theory as a testing ground. We
shall compare it to some existing results\cite{HelmutStephane}\cite{Luscher}
in the lattice regularisation. The Hamiltonian of the $\phi^4$ theory
is
\begin{equation}
H={1\over 2} \int d^3 x ( 1/2 ( ({\partial \phi \over \partial t})^2 +
  ({\partial \phi \over \partial \vec x})^2 + m_0^2 \phi^2) + {g_0\over 4!}
\phi^4)
\end{equation}
where $\phi(x)$ is the scalar field, $m_0$ and $g_0$ are the bare mass and
coupling constant respectively.

We set the scale $\Delta k $ to one and insert the discretised and quantised
solution

\begin{equation}
\phi(x)= \sum _{\vec k} {1\over \sqrt{(2\pi)^3}\sqrt{2 \omega(\vec k)}}
(a(\vec k) e^{-ik_\mu x^\mu}+a^\dagger (\vec k) e^{+ik_\mu x^\mu})
\end{equation}

of the kinetic field equations of motion back into the Hamiltonian, where
$a,a^\dagger$
are the usual annihilation and creation operators
 $[a^\dagger(\vec k),a(\vec l)]=\delta_{\vec k,\vec l}$ respectively and
$\omega(\vec k)=k_0=\sqrt{m_0 ^2 + \vec k ^2} $ is the kinetic energy of a
parton. This yields the discretised Hamiltonian that we diagonalised.
We denote the n-th eigenvalue of the Hamiltonian Matrix as $E_n$ and
the n-th physical mass as $M_n:=\sqrt{E_n ^2-\vec P ^2}$.
We did not normal order -- which
would not have been allowed for an interacting Hamiltonian -- but we
subtracted the vacuum energy. The latter is equivalent to the requirement
that the perturbative vacuum be annihilated by the Hamiltonian.
{\it Only} if the vacuum-energy $E_0$ is gauged to be zero can we use the
mass-shell condition
$M_n:=\sqrt{E_n ^2-\vec P ^2}$ to calculate the physical mass spectrum from the
energy spectrum. We do not have to take the field renormalisation constant
$Z_\phi$ into account, but
one problem arises. The critical line lies entirely in the region where
$m_0 ^2$ is negative and we cannot build up the Fock-space in terms of
partons with imaginary masses. Instead, we have chosen to build it up
in terms of partons with mass $m_K$ by splitting the $m_0^2\phi^2/2$ mass term
into a kinetic $m_K^2\phi^2/2$ and a dynamic $m_{int}^2\phi^2/2$ part where
$m_K ^2=m_0^2-m_{int}^2>0$ is the kinetic energy of the partons and
$m_{int}^2$ is a sufficiently large negative number that takes the
r{\^o}le of a coupling constant. We have presented our results for
the choice $m_K ^2= 0$. We have found, however,
that the critical behaviour was almost
independent of a specific choice of $m_K$ in
the $N=4$-sector, if only
$m_K ^2$ is reasonably small.

\section{Critical behaviour of the scalar $\phi^4$-theory}

We diagonalised the described Hamiltonian on a $7^3$ and a
$9^3$ lattice. To do this, Fock spaces
of only 6 resp. 21 states are needed(and even less if the
$Z_2$ symmetry is removed in the symmetric phase).
In order to compare our results to the work of L{\"u}scher and
Weisz\cite{Luscher}
and to the work of D. B{\'e}rub{\'e} et al.
\cite{HelmutStephane} we have to introduce the parameters
$\lambda$ and $\kappa$ which are related to $m_0$ and $\lambda$ by
$m_0^2=(1-2\lambda)/\kappa-8$ and
$g_0= 6{\lambda \over \kappa^2}$.
In Fig.1 the first of the three data columns of the renormalised masses
$m_R$
in terms of $\kappa$ as given by L{\"u}scher and Weisz is
shown and compared to our ground state masses $M_1$.
We have diagonalised the Hamiltonian
for $50\times 30$ different bare parameters $m_0 , g_0$ and fitted the results
to obtain the energy spectra in terms of $\lambda$ and $\kappa$.
It can be seen in Fig. 1 that the shape of the function is well
described even
on the $7^3$-lattice. For the $9^3$-lattice we are able to describe
most of the scaling region $M_1 \leq {1\over 2a}$.
The ground state mass $M_1(\kappa)$ inside the validity domain
$1/(2N)<<a M_1<<1$ of our model is seen to obey the scaling law
$M_1 \sim C_4 \tau^{1/2}|ln \tau|^{-1/6}$,
where $C_4$ is a constant\cite{Luscher}
and $\tau:=1-\kappa/\kappa_{crit}$ denotes the reduced
temperature, meaning that we have correctly described the critical exponent
$\nu=1/2$ plus
the logarithmic correction $|ln \tau |^{-1/6}$ in our {\it Hamiltonian}
approach.
It can also be seen that
the critical point is well reproduced. This is true also for curves
of different $\lambda$ close to the Gaussian fixed point
as it is shown in Table 1. where we listed the
quotients $\alpha_N := \kappa_{crit}/ \kappa^{(N)} _{crit}$;
the quantity $ \kappa^{(N)} _{crit}$ denotes the point where the
renormalised mass
$M_1$ becomes imaginary(It becomes real again in the asymmetric phase
if we break the $Z_2$ symmetry by choosing a non
zero field expectation value.). Table 1 shows that the quantity
$\kappa^{(N)} _{crit}$ is an excellent
estimator for the critical line.
In Fig.2 finally, the mass spectrum related to the first excited state,
$M_n/M_1$, is depicted for fixed $\lambda$ in terms of $\kappa$. It clearly
demonstrates the fact that the partons of the $\phi^4 _4$ theory interact by
repulsive forces. The first state above the
vacuum is a one particle state with negative parity and the positive parity
state above
consists of two particles.

\section{Distribution functions}

Since the $\phi^4$ theory describes partons with repulsive two particle
exchange forces and since it has a trivial, i.e. free continuum limit, its
distribution functions are not comparable to the QCD structure functions.
In the $\phi^4$ theory, a renormalisation group transformation(RGT)
decreases the lattice size as well as the marginal coupling constant.
In an asymptotically free theory such as QCD, the coupling constant
increases under such a transformation. This means a RGT
enforces particle creation
for QCD,
whereas particle creation is weakened for the scalar theory.
Since
the renormalisation group in the Wilson/Kadanoff sense
is used to make a complicated system numerically tractable by coarse-graining
it
into a theory with equivalent long distance behaviour, this means for the
$\phi^4$-theory that
the coarse-grained system has a strongly reduced particle content so that a
particle
cutoff may be applied after a few RGTs.
This can not be expected to hold for QCD where
a RGT actually increases the interaction. This effect slows down or may
even invert the net particle
number decimation of a RGT due to a reduction of momentum space
points
and this is
the problem we designed our method for.
The distribution function $f(x_B,Q)$ of finding some parton with
momentum fraction $x_B$ inside the hadron is determined by the
parton momentum distribution function $\tilde f(\vec p, \vec P)$ for
finding a parton with momentum $\vec p$ inside the hadron with momentum
$\vec P$. Since $Q$ is a dimensionful quantity, its scale is set by
the lattice spacing  $Q \sim  \displaystyle{1\over a(m_0,g_0)}$
and thus depends on
the bare parameters if one keeps the
renormalised mass fixed. The continuum limit $a\rightarrow 0$ corresponds
to the the limit towards arbitrarily high resolution ability. If one
keeps the renormalised mass and the renormalised coupling constant fixed,
then $Q$ is a function of the bare coupling constant $g_0$ and vice
versa -- invertibility of $Q(g_0)$ assumed. Hence $f(x_B,Q)$ is related by
the function $Q(g_0)$ to
the distribution function $\bar f(x_B,g_0)$ that only depends on dimensionless
parameters. Consequently, a calculation
of the distribution function along a renormalisation group trajectory
can be used to compute some peculiarities of the $Q$-dependence
 of the quark structure functions such as the increasing of the
peak at $x_B=0$ when
the resolution $Q$ is increased.
In order to demonstrate that the suggested method is capable of
describing them, we calculate the distribution functions of the
$\phi^3$ theory whose forces are attractive one particle exchange forces.
Although this theory is known to suffer from an
unstable vacuum, it serves well
for our purpose of demonstrating qualitatively
 that the singularity at $x_B=0$ and its change in terms of the resolution
$Q$ of experimentally measured
structure functions can be modelled in our approach
(The unstable vacuum of the $\phi^3$
theory does not allow us to calculate meaningful ground state masses that
are needed to specify renormalisation group trajectories and consequently
the exact relation between the resolution $Q$ and the bare coupling constant
$g_0$. But the
distribution function $\bar f(x_B,g_0)$ suffices for our purposes).
To increase the number of particles that can be modeled on our tiny
workstation, we have chosen to present the results of the
two dimensional $\phi^3 $ theory. The
results, however, are qualitatively alike in higher dimensions except
for the smaller particle number.
In Fig.3 we depict the
function $\bar f(x_B,g_0)$ to illustrate that our method is -- at least
in principle -- powerful enough to describe the known features
of the proton structure function.
This can {\it by no means} be taken for granted. Fock space methods that
have to apply a low particle number cutoff in addition to the phase space
cutoff are in principle
unable to show this feature\cite{DiplomArbeit}.
This is so because a system of n identical {\it observable} particles must have
an expectation value of $x_B$ around $<x_B>=1/n$ for symmetry reasons.
Furthermore, the behaviour
of the structure functions in terms of the resolution $Q$ is dominated not by
the power law rest
effects of the interaction between the partons but rather by the fact that a
higher resolution
power unveils {\it more} particles to the observer. With a low particle
number cutoff, the splitting functions appearing
in the Altarelli-Parisi equations would be meaningless, anyway.
For the $\phi^4$ theory, however, an additional particle cutoff is perfectly
feasible for describing the lowest-lying eigenstates of the Hamiltonian.

\section{Discussion}

We have shown that the suggested principle that only measurable quantities
be described works well for the scalar theory.
Even a $6\times 6$ Hamiltonian matrix is able
to correctly describe important features of the critical behaviour of the
$\phi^4$-theory. We are, however, aware of the fact that there is a huge
difference between
the scalar $\phi^4 _4$-theory which does not even possess a nontrivial
continuum limit and nonabilian gauge theories which do possess a continuum
limit with strongly bound particles and also
an additional topological structure.
But we have reasons for being optimistic.
Our method has been designed to sooth one
major nuisance of highly relativistic systems,
the uncontrolled particle
production.
This feature has not even been necessary for the $\phi^4$
theory where we could introduce a {\it modest} particle cutoff without
mutilating the results.
This is encouraging since
we did not aim at the scalar theory in the first place. The aim is the
calculation of structure functions, one of the outstanding
problems of particle physics,
directly from $QCD$ where high particle numbers are inevitable.
We hope that our results can be extended to
gauge theories and stress that the suggested method has the power to
significantly simplify the Kogut
Susskind Hamiltonian because it also works on a position space lattice.
This is important
since there is no $QCD$ Hamiltonian for momentum space available yet.
The Kogut-Susskind Hamiltonian being a position space
method is plagued by mirror fermions.
Our approach, taking only parton
momenta of the first octant in the brillouin zone,
{\it formally} eliminates them from the set of 'observable' partons,
which seems to be a remedy to this problem (We break the
mirror-symmetry by the choice of a specific
hadron momentum $\vec P$ different from $\vec 0$).
If this suffices to describe the axial anomaly remains to be seen.
We have to be very cautious here:
the Nielsen-Ninomiya no-go theorem\cite{NieNin}
has yet always resisted attempts to circumnavigate it.

\section{Acknowledgements}
One of the authors\footnote{Sic. Author{\it s}. improved version of this
paper to be
submitted as a letter.} (N.Scheu) wants to express his appreciation for
having been granted the AUFE fellowship from the DAAD (German Academic
Exchange Service)
which has made this Ph.D. project possible!

\newpage
\begin{flushleft}
{\bf Figure Caption}
\end{flushleft}
\begin{description}
\item[{Fig.1}]

The ground state mass $m_R(\lambda,\kappa)$ in lattice units ($a\equiv 1$) is
depicted as a function
of the lattice parameter
$\kappa$ for the the semi-analytical data of ref. \cite{Luscher}
(points) and for our data obtained on the $7^3 $-lattice(dashed line) and on
the $9^3 $-lattice(solid line). The parameter $\lambda $ is taken to be
$\lambda=0.00345739 $.

\item[{Fig.2}]

The lowest lying mass spectrum on the $9^3$ lattice
is depicted in terms of $\kappa$ for
$\lambda=0.00345739$. The ground state is set to one.

\item[{Fig.3}]

The distribution function $\bar f(x_B,g_0)$ of $\phi^3 _2$ in terms of the
momentum
fraction $x_B$ and the coupling constant $g_0$. The mass $m_0$ is set
to be $m_0=3 \Delta k$. The maximum particle number $N$ is $N=11$.

\end{description}

\newpage

\section{Table Caption}

 \begin{center}
 \begin{tabular}{|| c|c|c|c|c|c|c|c|c|c ||}\hline
$\lambda$&.001 &.005 &.01& .02&.03&.04&.07\\
\hline
$\kappa_{crit}$&
0.125202& 0.125991& 0.126968& 0.128604& 0.130096& 0.133096& 0.133825\\
\hline
$\alpha_3$&
1.01327& 1.01194& 1.01192& 1.01718& 1.0227 &1.02664&1.03075\\
\hline
$\alpha_4$&
1.00949& 1.00518& 1.0043& 1.00262& 1.00079&1.00068&0.95589\\
\hline
 \end{tabular}
 \end{center}

The semi-analytically determined critical points $\kappa_{crit}$ in terms
of the lattice parameter $\lambda$ and
the quotients $\alpha_N := \kappa_{crit}/ \kappa^{(N)} _{crit}$. The
quantity
$ \kappa^{(N)} _{crit}$ denotes the points we found where the renormalised mass
$M_1$ becomes imaginary. A tiny systematical error of about $1\%$ can be
expected to be present due to effects of our having interpolated
the function $M_1(m_0,g_0)$ on a finite (but large) number of parameter points.

\end{document}